\documentstyle[12pt,axodraw]{article}
\begin{document}
\centerline{hep-ph/0105177 \hfill UMD-PP-01-052}

\bigskip
\bigskip

\begin{center}
{\large \bf Models for Geometric CP Violation with
Extra Dimensions}
\vskip 1cm
Darwin Chang$^{a,b}$\footnote{e-mail: chang@phys.nthu.edu.tw},
Wai-Yee Keung$^b$\footnote{e-mail: keung@uic.edu},
and
Rabindra  N. Mohapatra$^c$\footnote{e-mail:rmohapat@physics.umd.edu}
\\
\vskip 1cm
{\small
$^a$ 
Physics Department, National Tsing-Hua University, 
Hsinchu, Taiwan, 230043, ROC\\
$^b$ Physics Department, University of Illinois at Chicago,  IL 60607-7059\\
$^c$ Department of Physics, University of Maryland, College Park, MD, 20742\\
}
\end{center}

\centerline{May, 2001}

\begin{abstract}
{In a recent paper, two of us (D.C. and R.N.M.) proposed a new way to
break CP symmetry geometrically using orbifold projections.  The mechanism
can be realized in the five dimensional brane bulk picture. In this
paper, we elaborate on this proposal and provide additional examples
of models of this type. We also note the phenomenological implications
of some of these models.}
\end{abstract}
\bigskip
\bigskip

\section{Introduction}
There are many puzzles and mysteries in the highly successful standard
model of electroweak interactions. One of the most prominent among them is
the origin of CP violation observed in the kaon system and more recently
perhaps, in the $B$-system. Another evidence for CP violation is in the
domain of cosmology where there is evidence for asymmetry between matter
and antimatter. 

To introduce CP violation into gauge theories, one starts with the
elementary field theoretic observation that complex couplings
generally imply CP violation. In the standard model, these complex
couplings\cite{KM} are introduced into the theory ``by hand'' and no
insight is
gained as to the origin of CP violation. A different way of introducing
CP violation is to have the original Lagrangian to be CP conserving but
to let the
vacuum state break CP\cite{lee}. This phenomenon is known as spontaneous
CP violation and it generically leads to a distinct picture for
the early universe, where CP symmetry may be restored. An intriguing early
suggestion in this context\cite{mz} is that the smallness of observed CP
violation may be due to the fact that CP violation arises as a quantum
effect. It is also worth remembering that a final resolution of the well
known strong CP problem of QCD may depend on our true understanding of the
origin of CP violation. 

In view of the significant role played by CP violation in a complete
picture of particle physics, it is important to seek different ways to
have a fundamental understanding of this phenomena. In a recent paper,
two of us (D. C. and R. N. M.)\cite{cm} proposed a new geometric way to
understand the origin of CP violation. The basic idea is to consider
a theory in five space time dimensions (4+1) with the standard
model  residing in a 3+1 dimensional brane and have a degenerate pair of
fields in the bulk (i.e. 4+1 dimensional space-time). It was then shown
that if the two members of the degenerate pair are given asymmetric
boundary conditions and their coupling to brane fields is suitably chosen,
then it can result in CP violation. This is a fundamentally different way
to introduce CP violation into particle physics from the ones known
to date. While the models we present are in five dimensions, the
mechanism can be easily generalized to arbitrary higher dimensions.

A key ingredient of this proposal is that there must be degeneracy of
states which allows for a generalized definition of CP transformation.
Often such degeneracy arises due to the very nature of five or higher
dimensional
space-time making the basic premise of these models very natural. For
instance, in five dimensions, a fermion is necessarily
four-component whereas in 4-dimensions, it can be 2-component; thus, the
two 2-component spinors of the five dimensional 4-component spinor could
be taken as the pair of particles. Another example is to consider
supersymmetry in five dimensions; in the bulk, it automatically becomes
an $N=2$ supersymmetry. If we consider a hypermultiplet of $N=2$, it has two
$N=1$ chiral multiplets. In the main body of the paper, we will give
examples of both types and display how CP violation really arises.

Clearly, this way of breaking CP symmetry has another aesthetically
appealing feature that now one can have an unified geometrical
understanding of all forces such  as gravitational, gauge as well as the
origin of CP violation, a dream of many physicists 
ever since Einstein's general theory of relativity provided a successful
description of gravitational forces. In this connection, it may be worth
noting that in recent literature, there are several other examples of
symmetry violation by geometrical effects e.g. parity\cite{perez}, weak
gauge symmetry \cite{hall} as well as the grand unification symmetries
\cite{kawamura}.

In Ref. \cite{cm}, we discussed several examples of models where geometric
CP violation arose from asymmetric boundary conditions in the bulk. In
this
paper we discuss several new examples and note some phenomenological
implications of these models. We also discuss the general issue of the
connection between
complex phases and CP violations and emphasize that the presence of a  
complex phase
in a theory does not necessarily mean CP violation, especially when
there are
degenerate particles. This basic observation is in some ways at the heart
of our new mechanism.

Note that we do not purport to have a complete explanation of the origin
of CP symmetry and its breaking.  We assume that CP symmetry arises  
automatically out of some higher energy theory in higher dimensions such
as string theory.  We also cannot explain why nature may choose to  
compactify itself in an orbifold construction which violates the CP  
symmetry that was otherwise endorsed by the higher energy theory.  These
are hard questions whose understanding would require a better
understanding of string theories themselves.  However,
it is very intriguing that given both possibilities, one can actually put
the low energy CP violation that has been observed on our brane world
into the context of a higher dimensional CP conserving world.

This paper is organized as follows: in Sec. 2 and 3, we present simple
examples
of models where a naive definition of CP transformation may suggest that
the theory is CP violating whereas a generalized definition clearly
demonstrates that
the theory is CP conserving. In Sec. 4, we give an example of a model that
relates CP and P violation using similar ideas. 
In Sec. 5, we
discuss another example where an apparently CP conserving theory turns out
to be exactly CP conserving once one considers generalized CP
transformtions. In Sec. 6, we show how
asymmetric boundary conditions in a compactified fifth dimensions can
lead to CP violation and show that in certain models bulk Lorentz
invariance forces the needed asymmetric boundary conditions. 
In Sec. 7, we discuss an extension of the standard
model, where asymmetric boundary conditions arise automatically leading
to the familiar Kobayashi-Maskawa model. Section 8 is devoted to a model
where both P and CP have geometric origin and in Sec. 9, we elaborate
on an example given in Ref.\cite{cm} where asymmetric boundary conditions
implied by the $N=2$ supersymmetry in the bulk leads to a specific CP
profile for MSSM. We discuss the phenomenological viability of this model.
In Sec. 10, we give our concluding remarks.

\bigskip

\section{Complex phases and CP violation}

It is generally believed that a field theory that has complex phases
leads to CP violating effects. The argument can be illustrated taking the
example of a real scalar field $\eta$ and a complex scalar field
$\phi$.  The relevant Lagrangian can be written as:
\begin{equation}
{\cal L}(\phi_1)~=~\lambda \eta \phi_1^* \phi_1 + m^2 \phi_1^* \phi_1 + (
h \eta \bar{e}_L e_R + m_e \bar{e}_L e_R + H.c.) \ .
\label{eq:eq1}
\end{equation}
Note that even without $\lambda$, the complex phase in $h$ cannot be
removed once the mass $m_e$ is made real.  This is reflected in the
electric dipole moment (EDM) of electron generated at one loop level, with
$\eta$ in the loop, which is proportional to $\hbox{Im}(h^2 m_e)$.  
There is an exception when $h$ happens to be pure imaginary.  In that
case the Lagrangian is CP conserving without $\lambda$, and $\eta$ can
be defined to be CP odd.  However with $\lambda$ included, the
theory is again CP violating because $\lambda$ is 
real by hermiticity, 
and  the $\lambda$ interaction dictates the $\eta$ has to be CP even, 
while the Yukawa term dictates the $\eta$ to be CP odd.  
This CP violation is reflected in the two loop contribution\cite{barr}
to the electric dipole moment\cite{ckp} as in Fig.1.  
The effect is proportional to $\lambda$ $\hbox{Im}(h)$.
\begin{center}
\begin{picture}(200,160)(0,0)
\DashCArc(100,80)(40,0,360){5}
\ArrowArcn(100,80)(40,50,40)
\ArrowArcn(100,80)(40,140,130)
\ArrowArcn(100,80)(40,280,260)

\Photon(100,120)(100,160){5}{5}
\Text(110,150)[lc]{$\downarrow$ $\gamma$}
\DashArrowLine(20,20)(60,60){5} 
\Photon(134,60)(180,20){3}{5}
\Text(35,40)[rc]{$\eta$} 

\Text(80,60)[cc]{$\phi_1$}
\Text(130,115)[lc]{$\phi_1$}

\Text(175,50)[lc]{$\gamma$}
\ArrowLine(0,20)(20,20) \Text(10,10)[cc]{$\ell$}
\ArrowLine(20,20)(180,20) \Text(100,10)[cc]{$\ell$}
\ArrowLine(180,20)(200,20) \Text(190,10)[cc]{$\ell$}

\end{picture}
\end{center}

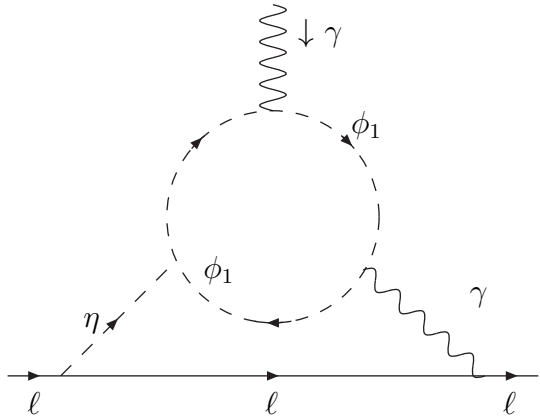
{\sl Fig.~1  The two-loop graph that contributes to 
the EDM  of the lepton $\ell$.  }

\bigskip

Let us now extend this model by including another complex scalar field
denoted by $\phi_2$, which is degenerate  in mass with $\phi_1$. Consider
the following Lagrangian:
\begin{equation}
{\cal L}~(\phi_1, \phi_2) = ~\lambda \eta (\phi_1^* \phi_1
-\phi^*_2\phi_2)+ m^2 (\phi_1^*
\phi_1 + \phi^*_2\phi_2)+ ( h \eta \bar{e}_L
e_R + m_e \bar{e}_L e_R + H.c.)  \ .
\label{eq:eq2}
\end{equation}
First, suppose $h$ is purely imaginary. Then, if we defined 
CP transformations as usual i.e. $\phi_i\rightarrow \phi^*_i$ under CP,
it would appear that the model violates CP.  However, if we define CP
transformation in a more general manner i.e. $\phi_1\rightarrow \phi^*_2$,
then the Lagrangian is CP conserving. The Yukawa interaction dictates the
$\eta$ to be CP odd.  With just one scalar $\phi_1$ as in Eq.(1), this CP
property is incompatible with the Higgs self-interaction.  However with
two degenerate $\phi_i$, in Eq.(2), it becomes possible to adjust the CP
property of $\phi_i$ such that $\eta$ remains CP odd. 
It is easy to see that the two loop
electric dipole diagram in Fig. 1, will now receive two
contributions from $\phi_{1,2}$ with opposite signs and equal
contributions leading to zero EDM of the electron.

It is also easy to see from this example that if the two scalar fields had
different masses, then there would be a CP violating contribution and one
will have a CP violating theory. This is a different way to introduce CP
violation into gauge theories. We will exploit this idea to propose new
kinds of models of CP violation, including those with extra dimensions  
and discuss their implications.

Another manifestation of how CP violation is connected to the degeneracy 
of the $\phi_{i}$ fields can be seen as follows. 
Suppose, we choose a potential for the CP-odd $\eta$ field as 
\begin{equation} 
V(\eta) = m^2_{\eta} \eta^2 + \lambda_{\eta}\eta^4 \ ,\end{equation}
with $m^2_{\eta}> 0$,  then $\langle\eta\rangle =0$. 
Thus vacuum also leaves CP as a good symmetry. Now if we take one loop
effects for the case where there is mass degeneracy,
the tadpole diagrams will cancel between $\phi_{1,2}$
keeping the $\langle\eta\rangle=0$ VEV stable under radiative corrections. 
However once the mass degeneracy between $\phi_{1,2}$ is removed,
there will be a nonvanishing tadpole contribution leading to a VEV of
the $\eta$ field and one will produce the breakdown of CP
invariance. Of course, one should note that while the VEV of $\eta$
breaks CP symmetry, CP is strictly speaking not broken
spontaneously.  The lost of degeneracy of the scalar masses already breaks CP symmetry softly.

This provides us a new way to relate CP violation with other new phenomena
in physics. For instance if the mass splitting between $\phi_{1}$ and
$\phi_2$ arose from parity violation, as we show in a subsequent section,
then parity violation become linked to CP violation providing
a new way to understand the origin of CP violation. Similarly, one
could relate this mass splitting to geometrical effects coming from extra
dimensions, leading to a geometrical origin of CP violation.

Note also that in the above examples, the coupling constants all can be
made real whenever a CP symmetry can be defined for the
Lagrangians.  However, one should note that having real coupling
constants is a  sufficient condition for CP symmetry but it is not
necessary.  In the appendix A, as well as in Sections 3 and 5, we provide
some examples in which CP is conserved even in a theory in which 
 there are some physical complex phases in the coupling constants.

\section{Fermionic example}
 
Next we shall consider models with fermions.  First, consider a
model with four left-handed chiral fermions $f_1, f_2, f_3, f_4$ of
charges $+, -, +, -$ respectively and a real scalar $\eta$.  
We define the C conjugated field of $f$ as  
$f^c=C^{\dagger} \gamma_0^T f^{\dagger T}$.
Thus
$$(f_2^T C f_1)^\dagger= 
f_1^{cT} C f_2^c= f_2^{cT} C f_1^c 
  \ , $$
$$
 (f_2^T C \sigma_{\mu\nu} f_1)^\dagger  =
 f_1^{cT} C \sigma_{\mu\nu} f^c_2  =- f_2^{cT} C \sigma_{\mu\nu} f^c_1 \ .$$ 
Let us study  the following Lagrangian 
\begin{equation}\begin{array}{lr}
{\cal L}_f=&\lambda    \eta(f_1^T    C f_2    - f_3^{cT} C f_4^c)
         +\lambda^*  \eta(f_1^{cT} C f_2^c  - f_3^{T}  C f_4)   \\
         &+\mu       (f_1^T    C f_2    + f_3^{cT} C f_4^c)
          +\mu^*     (f_1^{cT} C f_2^c  + f_3^{T}  C f_4)   \\
         &+\Delta    (f_1^T    C f_4    + f_3^{cT} C f_2^c)
          +\Delta^*  (f_1^{cT} C f_4^c  + f_3^{T}  C f_2)  \ . \end{array}
\end{equation}
We set up the system so that it respects the following CP symmetry,
\begin{equation}
f_1 \to f_3^c \ ,\quad f_2 \to f_4^c \ ,\quad \eta \to -\eta ,
\end{equation}
%%%%
that is, $\eta$ is CP-odd.   
One can make the mass parameters, 
$\mu$ and $\Delta$ real by changing the phase of the fermions.  
In this basis, it is clear that the complex phase of $\lambda$ is 
a physical parameter.  
However, it has nothing to do with CP symmetry.  It is interesting
to note that there is a one loop contribution (with $\eta$ boson in
the loop) to the following dipole-moment operator, $a f_2^T C
\sigma_{\mu\nu} f_1 $ as in Fig. 2, where the one loop coefficient $a$
is complex and is proportional to $\lambda^2\mu^* e$.  

\begin{center}
\begin{picture}(200,60)(0,15)
\Text(90,65)[tc]{$\eta$}
\DashArrowArc(90,20)(50,0,180){4}
\ArrowLine(0,20)(40,20)   \Text(20,10)[cc]{$f_1$}
\ArrowLine(90,20)(40,20)  \Text(65,10)[cc]{$f_2$}
\ArrowLine(90,20)(140,20) \Text(100,10)[cc]{$f_1$}
\ArrowLine(180,20)(120,20)\Text(160,10)[cc]{$f_2$}
\Text(40,10)[cc]{$\lambda$}
\Text(140,10)[cc]{$\lambda$}
\Text(90,25)[bc]{$\mu^*$}
\Text(120,25)[bc]{$e$}
\Line(87,17)(93,23) \Line(87,23)(93,17)
\Photon(120,20)(120,0){3}{3}
\end{picture}
\end{center}
%%%
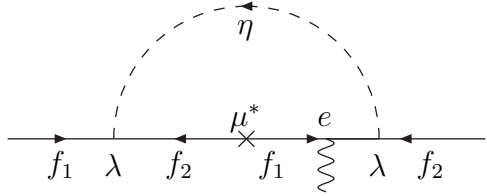
{\sl Fig.~2 The one-loop graph that contributes to electromagnetic
dipole moment.  The cross location denotes a mass insertion.  }
\noindent

There is also a similar diagram that uses the $\Delta$ mass insertion
instead of $\mu$, it gives rise to the operator 
$b f_4^T C \sigma_{\mu\nu} f_1 $ with the one loop coefficient $b$ which is real and is proportional to $\lambda \lambda^* \Delta e$.  

Similarly, there are corresponding diagrams with 
${f_1, f_2}$ replaced by $f_3, f_4$ which
give rise to $a^* f_4^T C \sigma_{\mu\nu} f_3$ and 
$ b f_2^T C \sigma_{\mu \nu} f_3$.

So, one has one loop contribution to the magnetic dipole moments
\begin{equation}
{\rm Re}(a) 
(f_2^T C \sigma_{\mu\nu} f_1 + f_1^{cT} C \sigma_{\mu\nu} f_2^c)
+ b 
( f_4^T C \sigma_{\mu\nu} f_1 + f_1^{cT} C \sigma_{\mu\nu} f_4^c)
+  (1,2) \leftrightarrow (3,4)
\ ,
\end{equation}
as well as the electric dipole moments 
\begin{equation}
i{\rm Im}(a)[(f_2^T C\sigma_{\mu\nu} f_1 - f_1^{cT}C \sigma_{\mu\nu}f_2^c)
            -(f_4^T C\sigma_{\mu\nu} f_3 - f_3^{cT}C \sigma_{\mu\nu}f_4^c)]
\ .\label{eq:fedm} \end{equation}
Note that in the limit that
$\Delta =0$, the Lagrangian has an $U(1)\times U(1)$ flavor symmetry.
In this limit, the $(f_1, f_2)$ forms a Dirac pair as usual and so is
$(f_3, f_4)$ pair and the two pairs are degenerate in mass.  In this
sense, the EDM operators above is completely identical to those of
ordinary fermion such as electron.  The main difference is that, due
to the degeneracy, these EDM's are not a direct signature of CP
violation here
because one can still define a conserved CP symmetry transforming the
EDM of Dirac pair $(f_1, f_2)$ into that of Dirac pair $(f_3, f_4)$. 

The situation is a lot analogous to that of ammonia molecule which has
double approximate degeneracy in its ground state with opposite parity
property.  The degeneracy is exact only when the tunneling between the
two degenerate states is ignored.  When tunneling is ignored, one can
show that both states have non-zero EDM but of opposite sign without any
breaking of CP symmetry.  It is also interesting to note that when
tunneling is included, the degeneracy is lifted, the original EDM
becomes just the transitional moment between the two non-degenerate
states.  The role of the tunneling is played by $\Delta$ in our field
theory example. 

When $\Delta$ is nonzero, the mass eigenstates are 
\begin{equation}
 F_\pm =(f_1\pm f_3)/\surd2 \ , G_\pm =(f_2\pm f_4)/\surd2 \ . \end{equation}
They have nondegenerate mass terms
\begin{equation}    
(\mu+\Delta)F_+^TC G_+   +  (\mu-\Delta)F_-^TC G_- + H.c. 
\ .
\end{equation}
Therefore $(F_+, G_+)$ and $(F_-, G_-)$ form two Dirac pairs
$H$ and $K$ of masses $\mu+\Delta$ and $\mu-\Delta$
respectively,
$$ H =  F_+ + G_+^c \ ,\quad K=F_- + G_-^c \ ,$$
$$ -\overline{H} H=- \overline{F_+}G_+^c - \overline{G_+^c}F_+ =
 G^T_+ C F_+  +H.c.  \ , $$
$$ -\overline{K} K=- \overline{F_-}G_-^c - \overline{G_-^c}F_- =
 G^T_- C F_-  +H.c. \ . $$
The original EDM operators proportional to $\hbox{Im}(a)$ in 
Eq.~(\ref{eq:fedm}) can now be written as 
\begin{eqnarray}
i({\rm Im} a)& [(  G_-^T C \sigma_{\mu\nu} F_+ 
                 - F_+^{cT} C \sigma_{\mu\nu} G_-^c)
             +  (G_+^T C \sigma_{\mu\nu} F_- 
                 - F_-^{cT} C \sigma_{\mu\nu} G_+^c) ] .
\end{eqnarray}
With the property, 
$ \gamma_5 H =  -F_+ + G_+^c$, $\gamma_5 K=-F_- + G_-^c$,
we rewrite the above expression as
the transitional electric dipole moment 
between two nondegenerate Dirac fields.  
$$
i({\rm Im} a)[ \overline K\sigma_{\mu\nu}\gamma_5 H+
              \overline H\sigma_{\mu\nu}\gamma_5 K   ] \ .
$$
%%%
It is also interesting to rewrite the original 
Yukawa coupling $\lambda$ in this new basis:
\begin{eqnarray}
{\cal L}_{Y}= \quad
     &\hbox{Re}(\lambda)&\eta [(F_+^T C G_-  +  F_-^T C G_+) + H.c.]     
\nonumber\\
 +\ i\ &\hbox{Im}(\lambda)&\eta [ (F_+^T C G_+  +  F_-^T C G_-) - H.c. ]\ ,
\nonumber\\
= & -\hbox{Re}(\lambda)&\eta (\bar KH+ \bar HK)
 +\hbox{Im}(\lambda)\eta (\bar H i\gamma_5 H + \bar K i\gamma_5  K) \ ,
\end{eqnarray}
reflecting the property that $\eta$ is a CP odd scalar.

One may wonder what happens to the two loop Barr-Zee type contributions
when a Yukawa coupling of $\eta$ to the electron is introduced, such as
$i h \bar{e} \gamma_5 e$ as in Fig.~3.  Each of $f_1, f_2$ or $f_3, f_4$
pairs contributes to the EDM of electron, however, the contributions come
with opposite signs such that they cancel overall, as required by CP
symmetry.

\begin{center}
\begin{picture}(200,160)(0,0)
{
\Oval(100,80)(30,50)(0) \Photon(100,110)(100,160){5}{5}
\Text(110,150)[lc]{$\downarrow$ $\gamma$}
\DashArrowLine(60,60)(20,20){5} 
\Photon(140,60)(180,20){5}{5}
\Text(35,40)[rc]{$\eta$}

\Line(97,53)(103,47) \Line(97,47)(103,53)  %% cross

\Text(130,115)[lc]{$f_2$} \LongArrow(135,102)(140,98)
\Text(60,115)[lc]{$f_2$}  \LongArrow(60,98)(65,102 )
\Text(120,45)[tc]{$f_2$}  \LongArrow(121,53)(117,51)

\Text(80,45)[tc]{$f_1$}   \LongArrow(80,53)(85,51 )

\Text(175,50)[lc]{$\gamma$}\LongArrow(160,55)(170,45) 

\ArrowLine(0,20)(20,20) \Text(10,10)[cc]{$\ell$}
\ArrowLine(20,20)(180,20) \Text(100,10)[cc]{$\ell$}
\ArrowLine(180,20)(200,20) \Text(190,10)[cc]{$\ell$}
}
\end{picture}
\end{center}

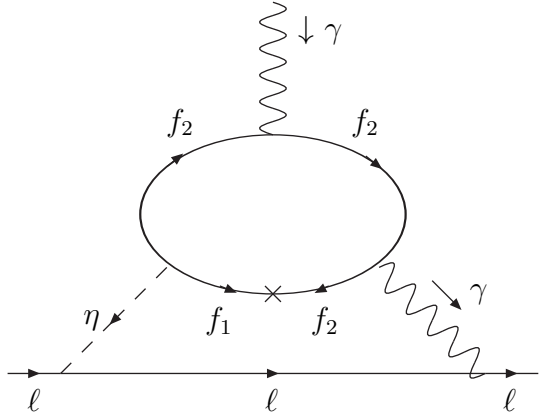
{\sl Fig.~3  The two-loop graph that contributes to EDM of lepton $\ell$.  The
cross location denotes a possible mass insertion.  }
\noindent

\bigskip

\section{Connecting P and CP violation}

In this section, we present a left-right symmetric extension of the
standard model where CP and P violation are connected to each other. The
strategy is to start with a theory which prior to spontaneous symmetry
breaking is both P and CP conserving. Using the analog of the bosonic
model discussed in section 2, we show that once parity is broken, it also
leads to CP violation.

We consider the usual left-right symmetric model\cite{lr} based on the
gauge group \\ 
$SU(2)_L \times SU(2)_R \times U(1)_{B-L}$ 
with fermion doublets 
$Q\equiv (u, d)$ and $\psi\equiv (\nu, e)$ assigned in a
left-right symmetric manner. For the symmetry breaking, we choose doublets
$\chi_L(2,1,1)$ and $\chi_R(1,2,1)$ and a bi-doublet 
$\phi(2,2,0)$. We
add to this model a real pseudo-scalar, C-even field $\eta$. 
We will assume that the Lagrangian prior to spontaneous symmetry breaking is invariant under the following P and CP symmetries.\\ 
Under P, we have 
\begin{eqnarray}
\eta   \leftrightarrow   -\eta           \ ,\quad
\phi   \leftrightarrow   {\phi}^{\dagger}\ ,\quad
\chi_L \leftrightarrow  \chi_R           \ ,\quad
Q_L    \leftrightarrow    {Q}_R          \ .
\label{LR}
\end{eqnarray}
Under charge conjugation C,
\begin{equation}
\eta   \leftrightarrow  \eta          \ ,\quad
\phi   \leftrightarrow  {\phi}^{T}    \ ,\quad
\chi_L \leftrightarrow  \chi^{*}_L    \ ,\quad
Q_L    \leftrightarrow   C\bar{Q_R}^T    \quad .
\end{equation}
The Yukawa interactions are 
\begin{equation}
f_{ij} \overline{Q_{Li}} \phi Q_{Rj} + 
g_{ij} \overline{Q_{Li}} \tilde{\phi} Q_{Rj} + H.c.
\end{equation}
The parity symmetry P implies the the coupling matrices, 
$f = f^\dagger$ and $g = g^\dagger$.  
The charge conjugation C  implies  $f = f^T$ and $g = g^T$.  
Therefore both coupling matrices are real and symmetric.

Parity symmetry is broken when the parameters of the
Higgs potential are chosen to be in a range such that 
$\langle\chi^0_R\rangle = v_R$ and
$\langle\chi_L  \rangle =0$. 
We will show that this also leads to CP violation. To see
this let us write down the relevant part of the CP conserving Higgs
potential:
\begin{eqnarray}
V^{\rm Higgs}  ~=~ -\mu_+^2(\chi^{\dagger}_L\chi_L +\chi^{\dagger}_R\chi_R) 
+ \lambda_+(\chi^{\dagger}_L\chi_L +\chi^{\dagger}_R\chi_R)^2
+ \lambda_-(\chi^{\dagger}_L\chi_L -\chi^{\dagger}_R\chi_R)^2\nonumber\\
+\lambda \eta (\chi^{\dagger}_L\chi_L -\chi^{\dagger}_R\chi_R)
+[m^2_{\phi}   \hbox{Det}\ \phi
+ i\lambda'\eta \hbox{Det}\ \phi + H.c.]   \ .
\end{eqnarray}
Note that in the limit of exact parity symmetry, $\chi_{L,R}$ have same
mass and the tadpole contributions to $\langle\eta\rangle$ cancel leading to
$\langle\eta\rangle=0$ and CP remains conserved. 
However as soon as parity is broken,
 the $\chi_R$ disappears from the spectrum and $\eta$ field has a nonzero
VEV and CP violation occurs. It gets transmitted to the quark and lepton
sector via the $\eta \hbox{Det} \phi$ coupling, which now gives complex VEV for
$\phi$. Thus CP and P violation get linked to each other. We do not
elaborate on the phenomenology of this model but defer it to a future
publication. We however note this way of relating P and CP violation is
different from the one in \cite{pati}.  We will also see how this scheme
can be more naturally imbedded in higher dimensional theory in Section 8.

\bigskip

\section{Duplicated standard model and CP violation}

Another example, which also illustrates the point that complex phases
need not necessarily imply CP violation is the model of
Ref.\cite{lav}. 
This model  uses the gauge group $SU(2)_A\times SU(2)_B\times
U(1)_Y$ and duplicates the field content of the standard model i.e.
$$q_L(2,1)_{1/3}\ ,
u_R(1,1)_{4/3}\ , 
d_R(1,1)_{-2/3}\ , 
\psi_L(2,1)_{-1}\ , 
e_R(1,1)_{-2}\ ,$$ 
as the usual standard model fermions where $(a,b)$ represents the 
$SU(2)_A\times SU(2)_B$ representations; the duplicate fermions are 
$$Q_L(1,2)_{1/3}\ ,
U_R(1,1)_{4/3}\ , 
D_R(1,1)_{-2/3}\ , 
\Psi_L(1,2)_{-1}\ , 
E_R(1,1)_{-2} \ . $$
There are two Higgs doublets $H_A$ and $H_B$, each a doublet under each group.
It is assumed that under CP, the lower case fields transform into the
upper case fields as:
\begin{eqnarray}
q_L &\leftrightarrow& \gamma_0C\bar{Q_L}^T\\ \nonumber
u_R &\leftrightarrow& \gamma_0C\bar{U_R}^T\\ \nonumber
d_R &\leftrightarrow& \gamma_0C\bar{D_R}^T\\ \nonumber
H_A &\leftrightarrow&  H^*_B  \quad .
\end{eqnarray} 
and similarly for other fields.
The CP invariant Yukawa coupling of the quarks can be written as:
\begin{eqnarray}
{\cal L}'_Y&=&\ 
\bar{q}_LH_A(h_d   d_R + h'_d D_R)+\bar{q}_L\tilde{H}_A(h_u u_R+h'_u U_R)
\\ \nonumber & & +
 \bar{Q}_L       H_B (h^*_d D_R+ {h'_d}^* d_R)
+\bar{Q}_L\tilde H_B (h^*_u U_R+ {h'_u}^* u_R)  
+ H.c. 
\end{eqnarray}
The coupling matrices $h_{u,d}, h'_{u,d}$ are all complex and yet the
theory is CP conserving. In fact as has been shown in Ref.\cite{lav},
CP violation arises only if 
$\langle H^0_A\rangle\neq \langle H^0_B\rangle$. 
If the two VEV's become equal, both sets of gauge
bosons have same mass and as a result, any linear combination of the
two sets of gauge bosons is also an eigenstate and this enables one to
get rid of all CP violating effects from the theory.  In Section 7, we 
will illustrate how this scheme can be easily imbedded in a higher  
dimensional theory so that CP violation arises geometrically from the
orbifold construction and results in an effective Standard KM model in
the brane.

\section{Orbifold boundary conditions and CP violation}

In this section, we show how the idea of
the previous section can be used to connect CP violation to the geometry
of space time. Consider for simplicity the complex scalar field model of
Section 2 and assume that electron and $\eta$ are brane fields whereas the
fields $\phi_{1,2}$ are bulk fields. 
Clearly the $\eta$-$\phi$ couplings in Eq.~(2)
involve the brane bulk coupling. Suppose we consider an $S_1/Z_2$ orbifold
where under $Z_2$ symmetry $y\rightarrow -y$. We then expect that under
the $Z_2$ symmetry ($R_P$), $\phi_i\rightarrow \pm \phi_i$. For even
$\phi$ fields the Fourier expansion will involve only the cosine modes
whereas for the odd fields, only the sine modes will appear. If we assume
that the brane is located at $y=0$, then on this brane the odd $\phi$
fields will vanish and the spectrum of the even and odd states will be
asymmetric in their mass spectrum. In particular, this will make the
$\eta-\phi$ coupling obviously CP violating. Coming to the electric dipole
moment of the electron in the toy model of section 2, the two loop diagrams
do not suffer from complete cancellation and one gets a non-zero EDM for
the electron.

In this example, one arbitrarily chooses the boundary conditions to get CP
violation and an objection could be raised that CP violation is in
some sense put in by hand although it is obviously connected to the
geometry of the fifth dimension. 
It is however possible to construct models, where
the asymmetric boundary conditions are dictated by kinematics of the
fifth dimensions.

As an example, consider a model where there is a fermion in the bulk;
its 5-dimensional kinetic energy can then be written in terms of the four
dimensional fields as 
\begin{equation}
 i\bar{\psi}\gamma^{\mu}\partial_{\mu}\psi 
+ (\bar{\psi}_L\partial_y\psi_R -\bar{\psi}_R\partial_y\psi_L)   \ .
\end{equation}
Due to the presence of the last
term, $Z_2$-invariance implies that $\psi_L$ and $\psi_R$ have opposite
$Z_2$ parity. As a result, if one of them has even Fourier components
i.e. cosines, the other field will necessarily have odd (sines) components
and therefore vanish on the brane at $y=0$. The asymmetry in spectrum
necessary for CP violation will then arise more naturally. Similarly,
if we have supersymmetry, then the bulk supersymmetry is $N=2$ type and
in terms of the $N=1$ supersymmetry, an $N=2$ hypermultiplet has two $N=1$
chiral superfields $(H,H^c)$. In the effective $N=1$ Lagrangian, there is a
term of the form $\int dy H\partial_y H^c$ term. This endows the $H$ and
$H^c$ fields with opposite $Z_2$ parity. This in turn leads to asymmetric
spectrum of fields and can be used to generate CP violation using our
idea.

Below we give examples of models where an effective CP violating theory
arises from the asymmetric boundary conditions described above.

\section{KM model from asymmetric orbifold boundary\\ conditions}

To see how the familiar CKM model can be obtained from orbifold
compactification, consider the model of the Sec. 5 in
5-dimensions, with the fifth dimension compactified on $S_1/(Z_2\times
Z'_2)$. In this
case, there are four kinds of states denoted by 
$$(+,+), (+,-), (-,+),(-,-) \ . $$
Except for the (+,+) states all other states have no zero modes
and are therefore not visible at low energies ($ E\ll R^{-1}$). Note that
the number of fermions doubles in the 5-dimensional model as compared to
the 4-dimensional one i.e. the states now are $(q_L, q_R)$ which are
$SU(2)_A$ doublets, $(u_R, u_L); (d_L, d_R)$ are singlets; similarly for
the leptons and the second set of quarks and leptons. We assign the
following $Z_2\times Z'_2$ quantum numbers to the quarks and leptons(see
table I):
\begin{center}
\begin{tabular}{|c||c|}\hline
Fields & $Z_2\times Z'_2$ quantum number \\ \hline\hline
$q_L, u_R,d_R, \psi_L, e_R, H_A$ & $(+,+)$\\ \hline
$q_R, u_L, d_L,\psi_R, e_L, H_B$ & $(-,-)$ \\ \hline
$Q_L, U_L, D_L, \Psi_L, E_L$     & $(+,-)$ \\ \hline
$Q_R, U_R, D_R, \Psi_R, E_R$     & $(-,+)$ \\ \hline
\end{tabular}
\end{center}
The CP invariant Yukawa coupling of the quarks in five dimension can be written as:
\begin{eqnarray}
{\cal L}'_Y&=&\ 
h_d \bar{q} H_A d + h_u \bar{q}\tilde{H}_A u +
h^*_d \bar{Q} H_B D + h^*_u \bar{Q} \tilde H_B U  
+ H.c. 
\end{eqnarray}
Note that $h'_d$ and $h'_u$ in Eq.(16) are not allowed by $Z_2 \times
Z'_2$ symmetry.  In the brane at $y=0$, only the fields with $(+, +)$
quantum numbers survive.   This leads to the familiar CKM model.  CP
symmetry disappears because of the asymmetry in the spectrum created 
by the orbifold construction.  The CP violating effect created by the
complex phases in $h_u$ and $h_d$ was originally cancelled by the similar
effects created by the CP conjugated states.  However the asymmetry in
the Kaluza Klein towers of the CP conjugated fermions destroy such
cancellation.

\vskip .5cm

\section{Common geometric origin of P and CP violation}

In this section, we use the left-right model of Sec. 4 to show both P
and CP violation can have a common geometric origin. We start with the
model of Sec. 4 in the brane and put a singlet neutrino 
$\nu^B$ in the bulk. The
brane field content is given by the
$SU(2)_L \times SU(2)_R \times U(1)_{B-L}$ gauge theory
with fermion doublets
$Q\equiv (u, d)$ and $\psi\equiv (\nu, e)$ assigned in a
left-right symmetric manner and Higgs fields
$\chi_L(2,1,1)$ and $\chi_R(1,2,1)$, a bi-doublet
$\phi(2,2,0)$ and the P-odd and CP-odd field $\eta$. The fields transform
under P and C as in Eq.~(\ref{LR}). 

The pure bulk part as well as the brane-bulk coupling terms are given by:
\begin{eqnarray}
i\bar{\nu}^B\gamma^{\mu}\partial_{\mu}\nu^B +
(\bar{\nu}^B_L\partial_y\nu^B_R-\bar{\nu}^B_R\partial_y\nu^B_L)
\nonumber\\ 
+\int dy \delta (y)
\left[\bar{\psi}_L\chi_L\nu^B +\bar{\psi}_R\chi_R\nu^B   +H.c.\right]  \ .
\end{eqnarray}
As discussed in Sec. 6, the $Z_2$ invariance of the bulk Lagrangian
implies that under $Z_2$ parity $\nu^B_L$ and $\nu^B_R$ have opposite
parity. Let us therefore assume that $\nu^B_L(x,-y) = \nu^B_L(x, y)$
whereas  $\nu^B_R(x,-y)= -\nu^B_R(x,y)$. This implies that for a brane
located at $y=0$, the $\nu^B_R$ field vanishes whereas the $\nu^B_L$
field appaears full strength. The effective brane theory therefore is
left-right asymmetric. As a result, the parity symmetry breaks. This
asymmetrizes the masses of $\chi_L$ and $\chi_R$. As already shown, under
this circumstance, the $\eta$ field will acquire a nonzero VEV and then
lead to CP violation. The CP violating phase is transmitted to the
fermions via the phases of the $\langle \phi \rangle$.

In order for the CP phase to manifest at low energies, one must have the
two standard model Higgs doublets in the bi-doublet $\phi$ survive at low
energies below the $W_R$ scale. If the $W_R$ scale in in the TeV range, as
possible in some extra dimensional models, then this does not require any
fine tuning. On the other hand, if the $W_R$ scale is high, some fine
tuning may be needed for the purpose.

\section{Profile of geometric CP violation in MSSM}

In this section, we apply this new mechanism to generate CP violation in
the minimal supersymmetric standard model (MSSM). For this purpose, we
start with the usual MSSM field content in the brane (i.e. $SU(2)_L\times
U(1)_Y$ gauge group and superfields $Q, L, u^c,d^c,e^c, H_u,
H_d$) augmented by the inclusion of a single superfield, which will be the
``messenger'' of CP violation. In the bulk we will now have $N=2$
supersymmetry. We will have two $N=2$ hypermultiplets in the brane, denoted
by its $N=1$ components $(H_1, H^c_1; H_2, H^c_2)$. Under CP symmetry, we
assume the MSSM fields to transform as usual i.e. $Q\rightarrow Q^c$, etc. 
The rest of the fields transform as follows:
\begin{eqnarray}
\eta\rightarrow -\eta^* \nonumber\\
H_1\rightarrow H^{c*}_2\nonumber\\
H_2\rightarrow {H^c}^*_1     \ .
\end{eqnarray}
We assume the theory prior to compactification to be CP symmetric so that
the only phase in the theory is in the coupling of the $\eta$ fields to
the bulk fields $H_{1,2}$:
\begin{equation}
W_{\eta}= \eta (\lambda H_1H_2-\lambda^*H^c_1H^c_2) + M_1\eta^2
+M_2(H_1H_2 + H^c_1H^c_2) \ ,
\end{equation}
where $M_{1,2}$ are masses expected to be of order of the fundamental
scale of the theory. It is possible to have a theory where the mass
parameters $M_{1,2}$ and the familiar $\mu$-term could arise from 
a K\"ahler potential of the form\cite{giu}
\begin{eqnarray}
S_K~=~\int d^4\theta \frac{S^{\dagger}}{M_{P\ell}}\left[ (H_1H_2 +
H^c_1H^c_2) +\beta H_uH_d\right]  \ .
\end{eqnarray}

Now note that since the bulk kinetic energy leads to
a term of the form\cite{hall} $H\partial_yH^c$, the required condition for
CP violation i.e. $H$ and $H^c$ have opposite $Z_2$ parity is
automatically satisfied and CP violation in the brane will ensue rather
naturally due to asymmetric spectrum of the bulk fields.

To see the profile of CP violation, let us write down the superpotential
in the brane involving the $\eta$ fields (the usual MSSM superpotential
terms involving the MSSM fields are omitted for simplicity). To
incorporate supersymmetry breaking, we have the usual hidden sector
mechanisms in mind. We will use a singlet field $S$ to implement the SUSY
breaking by choosing $\langle F_S \rangle = M^2
\approx (10^{11})$ GeV$^2$.
\begin{equation}
 W_{brane}=(i\eta + M_{wk})(a + b {S \over M_{P\ell}}) H_u H_d   \ .
\end{equation}
We have not written terms that are suppressed by higher powers of
$M_{P\ell}$ since their effect on CP violation is negligible.

CP violation in the MSSM arises when the field $\eta$ acquires a nonzero
VEV via the tadpole diagrams involving $H_{1,2}$ fields. In the
supersymmetric limit, due to the nonrenormalization theorem of
supersymmetry, $\langle\eta\rangle=0$. It is then easy to see that 
$\langle\eta\rangle\simeq {M_{susy}}/({16\pi^2})$ if the
parameter $m_{\eta}$ is in the TeV scale.
This leads to a profile of MSSM CP violation
where the only CP violating terms are the $\mu$ and the $B\mu$ terms.
All other CP violating phases in this model are extremely tiny due to
Planck mass suppression. For instance, to get CP violation in the squark
masses, one will have to write operators of type 
%%%
$$\int d^4\theta \frac{\eta SS^{\dagger}}{M^3_{P\ell}}Q^{\dagger}Q
\ .$$ 
%%%
After the $\eta$ field acquires a VEV, the resulting phase is of order
$10^{-16}$, which is clearly too small.

 By redefining one of the Higgs superfields, we can make the $B\mu$ term
real. So the only complex parameter in the theory is the $\mu$ term.
Furthermore, the CP phase is naturally of order $10^{-2}$ due to the
presence of the factor $16\pi^2$ above. 
It could of course be larger if the mass parameters in the theory
are adjusted (say somewhere between) 0.1 to 0.001, without going beyond 
usual naturalness requirements. There is no CP phase of the usual KM
type in this model.

Let us briefly comment on whether such a model can explain observed CP
violation in the kaon system. The only complex phase in the theory appears
to be in the term 
%%%
$$ |\mu| e^{i\alpha} h_{d,ij}  \tilde{Q}_iH^*_u \tilde{d^c_j} 
 + |\mu| e^{i\alpha} h_{u,ij}  \tilde{Q}_iH^*_d \tilde{u^c_j} \ , $$ 
which gives rise to the 
squark $LR$ mixing,
\begin{eqnarray}
\delta^d_{LR,ij} = \frac{((A_d + |\mu| e^{i\alpha}
\tan\beta) m^d)_{ij}}{M^2_{SUSY}}
\ ,
\end{eqnarray}
where $m^d$ is the down quark mass matrix, and $A_d$ is the trilinear
soft supersymmetry breaking coupling matrix which is in general not the
identity matrix at the weak scale, even if it may be the identical matrix
at the supersymmetry breaking scale. 
It has been shown in Ref.\cite{mura} that for $m_{\tilde{Q}}\sim
m_{\tilde{G}}\simeq 500$ GeV, the constraints from $\Delta m_K$,
$\varepsilon$  are respectively
(for small phase $\alpha$) 
$$\left(\hbox{Re} \delta^d_{LR,12}\right) \leq 4.4 \times 10^{-3} \ ,\quad
2 \left(\hbox{Re} \delta^d_{LR, 12}\right)
  \left(\hbox{Im} \delta^d_{LR, 12}\right)\leq 3.5 \times 10^{-4} \ .$$ 
If we saturate these values in our model, we then get
$|\varepsilon'/\varepsilon|$ to be $1.4 \times 10^{-3}$ which is good
agreement with experiments. The value of the electric dipole moment of
neutron can be made smaller than the experimental limit $11 \times
10^{-26}$ e$\cdot$cm if $|\hbox{Im} \delta^d_{LR,11}| \leq 3.0 \times
10^{-6}$ for  $m_{\tilde{Q}}\sim m_{\tilde{G}}\simeq 500$ GeV; and the
effective $\sin 2\beta$ parameter for $B$-decays to of order)
0.1 or less.
These two predictions could be used to test this particular realization
of our idea.

\section{Conclusion}

In this paper, we have elaborated on a novel mechanism for breaking CP
symmetry, suggested recently by two of the authors where the compactified
geometry of the fifth dimension played a crucial role. For this reason it
was called geometric CP violation. The
essential idea is that the asymmetrization of the spectrum by orbifold
conditions can lead to CP violating effects. Note that this is very
different from many recent papers\cite{recent} on CP violation in models
with extra dimensions in which CP violation is put into either a Higgs
VEV on some other brane or a susy breaking VEV's.  In our case, the
mechanism is genuinely geometrical in nature.  In this paper, we present
several new realistic models that provide realizations of this idea and
clarify the role of generalized CP transformations in implementing it.
While we cannot explain how the CP symmetry arises in the fundamental
higher dimensional theories, one application of this idea could be in the
domain of string theories where, it is likely that the very large gauge
symmetry which is the essense of the theory will be so
constraining that there will be no room for a CP violation at the
fundamental level. In that case, as pointed out here, one can imagine
that the CP violation observed in the low energy theory is due to the
particular compactification of the extra dimensions that somehow is
favored by the dynamics of the fundamental theory.

We wish to thank Hsin-Chia Cheng, We-Fu Chang and Lincoln Wolfenstein for
discussions and DC
wishes to thank Theory Group of University of  Maryland for hospitality
while this work was developed. 
The work of R. N. M. is supported by the NSF grant No. PHY-9802551 and that 
of D. C. and W. Y. K supported in parts by National Science Council of
R.O.C. and by U.S. Department of Energy (Grant No. DE-FG02-84ER40173).

\section*{Appendix A}
%%%
In this appendix, we discuss the connection between complex phases in a
Lagrangian and existence of CP violation. While it is generally true that
any phase that cannot be removed by redefinition of complex fields in the
theory is a physical phase and can lead to CP violation, we have found
examples where a physical phase does not lead to CP violation.
A simple example is provided in the following Lagrangian with two
complex scalar fields $\phi_{1,2}$ and one real scalar field $\eta$:
\begin{equation}
(\lambda \eta  \phi_1^* \phi_2 + \delta m^2  \phi_1^* \phi_2 +  H.c. ) 
+ m_1^2 \phi_1^* \phi_1  + m_2^2 \phi_2^*\phi_2 
 \ , 
\end{equation}
where $\lambda$ is a trilinear coupling. Note that by redefinition of the phases of the field $\phi_{1,2}$, 
one can make either $\delta m^2$ or $\lambda$ real but not both.  The complex phase of $\lambda$ in 
the basis in which $\Delta m_2$ is real is clearly physical.  However, it has nothing to do with CP violation.  
This can be seen by going to the mass eigenstate basis 
\begin{eqnarray}
\Phi_1 &=& +\phi_1 \cos \theta + \phi_2\sin \theta   \ , \\
\Phi_2 &=& -\phi_1 \sin \theta + \phi_2\cos \theta   \ .
\end{eqnarray}
In that case the most general Lagrangian can be written as 
\begin{eqnarray}
\lambda' \eta \Phi_1^* \Phi_2 + \lambda'^* \eta \Phi_2^* \Phi_1
+ \lambda_1 \eta \Phi_1^* \Phi_1 + \lambda_2 \eta \Phi_2^* \Phi_2
+ m_1^2 \Phi_1^* \Phi_1  + m_2^2 \Phi_2^*\Phi_2 .
\label{eq:eq3}
\end{eqnarray}
The phase of $\lambda'^*$ is unphysical and can be removed by redefining
the phase of, say, $\Phi_2$.  Therefore in this basis, all the couplings
are real and the theory has an obvious CP symmetry.  Note that the
corresponding couplings in Eq.~(\ref{eq:eq2}) can be identify as 
$\lambda_1 = - \lambda_2 = 2 \hbox{Re}(\lambda)$ and $\lambda' = i
\hbox{Im}(\lambda)$.  The phase of $\lambda$ ${\it is}$ a physical
parameter, but has nothing to do with CP violation.  
One can find a CP symmetry for Eq.~(2) defined as 
$\Phi_1 \rightarrow \Phi_1^*$, 
$\Phi_2 \rightarrow - \Phi_2^*$ and $\eta$ being CP even.  Note however
if one adds the coupling to the lepton in the form $i h \bar{e} \gamma_5
e$ plus electron mass, then the theory become CP violating because the
new coupling, $h$, forced the $\eta$ to be CP odd instead.  It will be
reflected in a two loop contribution to electron EDM as in Fig.1 and the
contribution is proportional to 
$\lambda_1 \hbox{Im}(h)$ or $\lambda_2 \hbox{Im}(h)$
depending on what is running in the inner loop.  CP would be conserved
if $\lambda_1 = \lambda_2 = 0$.

\section*{Appendix B}
%%%
In this appendix, we discuss a few elementary properties of fermions in
five dimension. The ``Lorentz'' group in 5--dimensions is SO(4,1) and its
algebra is specified by five $\gamma$ matrices $\gamma_{0,1,2,3,5}$. We
will choose the metric to be $(+, -,-,-,-)$. We choose a basis in which
the $\gamma$ matrices are given by (we use $i,j=1,2,3$ to be the known space
indices; 0 stands for the time index and 5 for the 5th component; often in
the text, we use $y$ to denote the extra space index i.e. $y=x_5$).
\begin{eqnarray}
\gamma_i = \left(\begin{array}{cc} 0 & \sigma_i \\ -\sigma_i & 0
\end{array}\right)\ ;\quad
\gamma_0 = \left(\begin{array}{cc} 0 & {\bf 1} \\ {\bf 1} & 0
\end{array}\right)\ ;\quad
\gamma_5 = \left(\begin{array}{cc} {\bf 1} & 0 \\ 0 & {\bf -1}
\end{array}\right)\ .
\end{eqnarray}

Each 4-dimensional Weyl spinor is a two-component object.
However, \\the 5-dimensional spinor $\Psi$ is necessarily 
a four-component one. 
Using $\gamma_5$, we can obtain from $\Psi$ 
the chiral 4-dimensional spinors by the usual formulas i.e.
${1\over2}(1\pm \gamma_5)\Psi\equiv \Psi_{L,R}$.

One can define two kinds of parity transformations in this case: the usual
P-parity under which ${\bf x}\rightarrow -{\bf x}$ and a fifth dimensional
parity (bulk parity or $Z_2$ parity) under which $y\rightarrow -y$. As
already mentioned in the text, invariance of the 5-dimensional kinetic
energy term under bulk parity implies that under this $\Psi_L$ and
$\Psi_R$ transform oppositely. As a result, the Dirac mass term
$\bar{\Psi}\Psi$ is not invariant under the bulk parity. However, the
assignment of the absolute bulk parity is arbitrary. It therefore follows
that if there are more than one bulk fermion, it is possible to assign
bulk parities in such a way that one has mass terms involving the bulk
fermions.

Turning to charge conjugation $C$, in the 4-dimensional case, it is usual
to define it by the relation $C\gamma_{\mu} C^{-1} = -\gamma^T_{\mu}$. As
a result, one can obtain $C=\gamma_2\gamma_0$. The $C$ however commutes
with $\gamma_5$. As a result, in five dimensions, one cannot use this
definition of $C$ and maintain 5-dimensional Lorentz invariance. The
fifth component of the kinetic energy is not C invariant. 
 One way
to maintain this definition of charge conjugation in the five dimensional
case is to simultaneously to transform the $y \rightarrow
-y$\cite{Thirring}.  This definition of C forbids the appearance of the
the mass term for bulk fermions. In this case, one can forbid a Dirac mass
term by $Z_2$ invariance.

Another way is to
define  $C= \gamma_2\gamma_0\gamma_5$, which satisfies the property
$C\gamma_aC^{-1}=\gamma^T_a$ where $a=0,1,2,3,5$. Note that 5-dimensional
C-invariance also forbids the Dirac mass term involving $\Psi$ but not
the Majorana mass.

\end{document}